# Minimizing the Latency of Quantum Circuits during Mapping to the Ion-Trap Circuit Fabric

Mohammad Javad Dousti and Massoud Pedram
University of Southern California
Department of Electrical Engineering
Los Angeles, CA 90089, U.S.A.
{dousti, pedram}@usc.edu

*Abstract*— **Quantum computers are exponentially faster than their classical counterparts in terms of solving some specific, but important problems. The biggest challenge in realizing a quantum computing system is the environmental noise. One way to decrease the effect of noise (and hence, reduce the overhead of building fault tolerant quantum circuits) is to reduce the latency of the quantum circuit that runs on a quantum circuit. In this paper, a novel algorithm is presented for scheduling, placement, and routing of a quantum algorithm, which is to be realized on a target quantum circuit fabric technology. This algorithm, and the accompanying software tool, advances state-of-the-art in quantum CAD methodologies and methods while considering key characteristics and constraints of the ion-trap quantum circuit fabric. Experimental results show that the presented tool improves results of the previous tool by about 41%.**

## I. INTRODUCTION

Quantum computers are able to solve some NP-intermediate problems in polynomial time, for which no deterministic, polynomial-time algorithm is known on classical computers. Quantum computers, which function based on the laws of quantum physics, are intrinsically different from classical computers. They use quantum bits or *qubits* instead of classical bits to represent information. A qubit is like a classical bit i.e., it can be set to either 0 or 1, but in addition, it can also present any superposition of 0 and 1.

Similar to traditional computers, a CAD flow is required for quantum computers not only to streamline the design process, but also to enable design of large and complex quantum circuits. Figure 1 shows a typical CAD flow for quantum computers. The gray blocks denote optimization tasks that benefit from design automation.

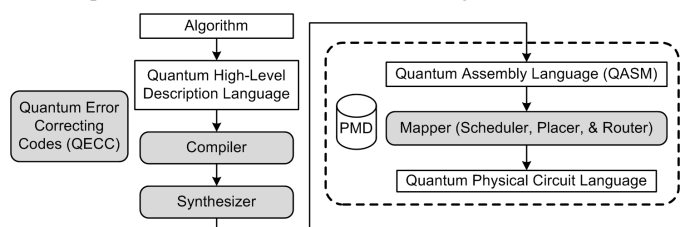

Figure 1. A typical Quantum CAD flow for ion-trap quantum circuit fabric. PMD refers to the physical machine description, which is different for different quantum circuit implementation technologies. Blocks in the dashed box identify the focus of this paper.

This paper focuses on the *mapper*. This tool maps the output of synthesizer, which is stored in a *Quantum Assembly Language* (QASM) file, to a given quantum circuit fabric. This step comprises of scheduling QASM instructions, mapping them onto the target *Physical Machine* (previously referred to as the quantum circuit fabric) and issuing the needed commands to control the underlying physical machine to run the desired operation in the correct order. Several efforts in this direction have been reported in the literature which will be reviewed at the end of this section.

A key challenge in building quantum computers that are big enough to solve real-life problems is the environmental noise - more precisely, loss of quantum information due to unwanted interactions between the quantum system and its environment. The circuit error should remain below a certain error threshold so the result has enough fidelity. *Error threshold* can be calculated as a function of underlying physical technology, employed quantum error correction (QECC), and fault-tolerant quantum control techniques.

The synthesis tool is responsible for adding QECC to increase the error threshold. Unfortunately, it cannot determine the circuit error before mapping, since it is unaware of total latency of the circuit. Mapping circuit to a fabric can change the latency of the circuit because of scheduling, placement, and routing steps [1]. Error analysis after circuit is being mapped can be done to determine the error and redo the synthesis step if the error threshold constraint is violated. More precisely, if the error threshold is not enough and the circuit takes longer time than expected, the circuit needs more encoding to improve the error threshold. In this work we focus on minimizing the total latency of the circuit to minimize the error in the circuit. A CAD tool, called *Quantum mapper based on Scheduling, Placement, and Routing* or QSPR, is developed to perform this task automatically. QSPR is not the first tool of its kind, but it improves the performance of existing tools dramatically. The rest of this section gives a survey of previous quantum CAD tools and summarizes the contributions and unique features of QSPR compared to prior art.

QUALE is a set of tools for the design and analysis of microarchitectures for ion-trap quantum computers [2]. One of QUALE's main tools does mapping of quantum instructions. The tool first creates a *quantum instruction dependency graph* (QIDG) to capture the dependency of instructions. The QIDG is traversed backward to schedule the instructions in an as late as possible (ALAP) manner. The tool models the quantum architecture layout (what we have termed quantum circuit fabric and is a finite-size regular grid) as a graph and the routing is done on this graph. A revised version of Pathfinder [3] is subsequently used for routing and dealing with resource contentions.

QUALE's placement technique is called *center placement*. In this method, qubits are placed in the free traps closest to the center of fabric. Although this method seems to be inefficient, since it places qubits next to each other, the delay cost of routing qubits is decreased. The key disadvantage of the center placement is that it is independent of the structure of the given QIDG. Hence, two qubits that have a lot of interactions may be placed far from each other. This increases the routing cost of bringing the two qubits together to perform the required gate level operations.

QPOS uses a similar flow as QUALE, but distinguishes between the source and destination operands of a two-qubit instruction during the routing step [4]. More precisely, the destination qubit is fixed in some trap while the source qubit is moved to reach the destination. Extraction of instructions from the QIDG is done in an as soon as possible (ASAP) fashion based on some priority function. In particular, the initial priority of an instruction is set to be the number of instructions that depend on it. QPOS extracts a routing path for each of the ready-to-issue instructions. If there are any overlaps among these paths, QPOS selects an instruction to execute (i.e., the source

This research was sponsored in part by a grant from the National Science Foundation.

qubit moved toward the destination site) based on the following criteria: 1) highest initial priority, 2) lowest amount of congestion that is going to be introduced by using the path, and 3) shortest path length. Finally, QPOS maps these paths to the quantum circuit fabric and uses a deadlock prevention algorithm to prohibit qubits to locate in a position that further movement is impossible. Reference [5] tweaks the QPOS by assigning the initial priority of instructions as the total delay of dependent instructions.

QSPR improves the state-of-the-art with respect to the aforesaid in the following important ways:

- It utilizes recent advances in the implementation of the ion-trap technology in terms of multiplexing ions in channels and traps and thus reduces the total execution latency of the algorithm that is being mapped to the quantum circuit fabric.
- It offers a more optimized global placement of the qubits on the quantum circuit fabric.
- It improves the routing algorithm by simultaneously moving both the source (control) and destination (target) qubits so as to minimize the movement delay. In addition, it considers turn delays when finding the best route between source and destination qubits.

The remainder of this paper is organized as follows. Section II describes the quantum circuit and ion-trap technology basics. Section III explains the scheduling method used. In section IV, the placement and routing methodologies are described. In section V, the experimental results based on QSPR are presented. The results are compared with QUALE, which was the only tool available for the public. Section VI concludes the paper.

## II. CIRCUIT REPRESENTATION AND ION-TRAP TECHNOLOGY

### A. Quantum Circuit and Its Representation

Figure 2 shows an encoding circuit for cyclic quantum error-correction. Note that in this circuit, four ancillary qubits are initialized to 0. Each qubit takes part in several one- and two-qubit operations. The 5 physical qubits at the output of the circuit represent a bundled logical qubit capturing and protecting (within the fault tolerance limits) the single input physical qubit. Figure 3 shows the QASM description of this encoding circuit.

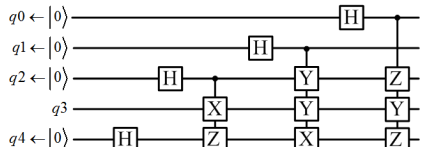

Figure 2. [[5,1,3]] encoding circuit; for cyclic quantum error-correcting [6]

| # | Instruction | # | Instruction | # | Instruction |
|---|---|---|---|---|---|
| 1 | QUBIT q0,0 | 6 | H q0 | 12 | C-Y q2,q1 |
| 2 | QUBIT q1,0 | 7 | H q1 | 13 | C-Y q3,q1 |
| 3 | QUBIT q2,0 | 8 | H q2 | 14 | C-X q4,q1 |
| 4 | QUBIT q3 | 9 | H q4 | 15 | C-Z q2,q0 |
| 5 | QUBIT q4,0 | 10 | C-X q3,q2 | 17 | C-Y q3,q0 |
|   |   | 11 | C-Z q4,q2 | 18 | C-Z q4,q0 |

Figure 3. QASM program description of the [[5,1,3]] encoding circuit.

### B. Ion-Trap Technology

Ion-trap technology is the most promising technology for implementing quantum circuits to date. Hence, it is selected as the underlying technology of QSPR. This technology provides a quantum circuit fabric with the following attributes [2][4][5][7]: *Qubits* are realized by ions. The number of available qubits must be provided in the circuit fabric specification; *Channels* act as wires for qubits. They allow qubits to travel inside them. Channels have two types: *horizontal channels* and *vertical channels*; Vertical and horizontal channels are connected together through the *junctions*. When a qubit wants to travel from a horizontal channel into a vertical channel or vice versa, it needs to make a *turn* at some junction; *Traps* are sites where quantum operations are performed. They are connected to the channels, so qubits can reach them by traveling through the channels. A trap is called *free*, if no qubit occupies it. In a 1-qubit operation, only one qubit resides in a trap. In a 2-qubit operation, two qubits inhabit a trap.

Figure 4 shows a 45×85 ion-trap fabric model, which was defined and released as a part of QUALE package [2]. A junction or a trap occupies a unit square, whereas a channel occupies one or several squares. In the latter case, the squares are aligned in a line. White spaces in this figure represent empty locations on the fabric.

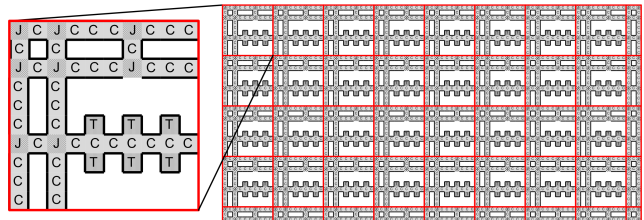

Figure 4. A 45×85 ion trap fabric. J represents a junction, C represents a channel, and T represents a trap.

Qubits travels through the channels and junctions to reach a trap in order to interact with each other. Two kinds of relocations are possible for qubits:

1. **Move:** This relocation happens when a qubit changes its place by one cell without changing its previous direction.
2. **Turn:** When a qubit wants to change its movement direction, first it needs to *turn* and then move to the new location. Note that a turn is a form of relocation that happens very slowly. A turn typically takes 5 to 30 times longer than a move [1].

References [8][9] have shown that the separation of two ions in a trap or channel is possible. Recently, reference [10] demonstrated that the proposed approaches in [8][9] can be integrated to build quantum computers with large number of ions. Based on this development, we introduce the notion of a *channel capacity* as the maximum number of qubits that can be concurrently sent through a channel. This is a technology-specific parameter. In this paper, we set the channel capacity to two. Note that junctions are designed such that we can route up to two qubits from any incoming channel(s) to any outgoing channel(s.) For example, we can route two qubits from the left to the top channel or one from the left to the top channel and another one from the bottom to the right channel. A channel or junction is called *free* if it has residual free capacity.

## III. INSTRUCTION SCHEDULING

The scheduling problem we must solve belongs to the class of Minimum-Latency Resource-Constrained (MLRC) scheduling problems. Resource constraints are defined by the channel and junction capacity limitations as specified above. Unfortunately, there is a complication here that differentiates between our scheduling problem and that found in standard high level synthesis tools for ASIC design. The complication is as follows. The delay of an instruction found in the QIDG of the quantum computational task (QASM program) is calculated as follows:

$$Instruction\ Delay = T_{gate} + T_{routing} + T_{congestion} \quad (1)$$

where $T_{gate}$ is the delay of required quantum operation, which is a known and fixed technology-dependent parameter; $T_{routing}$ is the move and turn delays of the operand qubits on the fabric to reach the target trap; and $T_{congestion}$ is the time an instruction waits in order to get access to a free channel or junction for routing. The last two delay values become known only after placement and routing steps are completed.

A complete solution to this problem requires an initial placement of qubits and simultaneous scheduling of instructions and routing of the quantum operands. Our approach schedules new instruction(s) after routing of each issued instruction. A priority value is defined for each unscheduled ready instruction as a linear combination of the number of unscheduled operations that depend on it plus the length of the longest path delay from that instruction to the end node of the QIDG. Higher priority instructions are scheduled first.

## IV. LAYOUT

### A. Placement of All Qubits

The placement of qubits will greatly affect the routing and congestion costs of any resulting solution. The goal of the placement step is to minimize the total computation latency of the scheduled QIDG. It is the sum of instruction latencies for the worst case (longest delay) path in the QIDG. The problem is hard, so we resort to heuristics to solve it.

We make use of the fact that the quantum computations are fully reversible, that is, given a QIDG, there is *one-to-one correspondence* between input and output qubits and that the inputs can be produced from the outputs by reversing the directions of all edges in the QIDG and performing the inverse gate level operations everywhere. We call QIDG with inverse gate-level operations for its nodes and reversed edges, the *uncompute* QIDG or simply UIDG. We can thus define a *forward latency* when a QIDG is executed according to a given total order (schedule) on a given circuit fabric; similarly we can define a *backward latency* for the case that the corresponding UIDG is executed according to the reverse scheduling order. The forward and backward latencies are the same if we ignore the routing and congestion delays. Otherwise, they may be different from one another.

The proposed placement algorithm works as follows. We start with an arbitrary *center placement* for the input qubits (see discussion about QUALE's placer in the section I.) Let's denote this initial placement as $P_1$. The schedule is known based on the previous step; let's call this total ordering on all instructions (gate-level operations) S. The reversed schedule (obtained by simply reversing the total ordering of instructions) is denoted by S*. Given $P_1$, we can then execute each instruction in the QIDG according to S. Details of the router are explained below. As a result, we obtain a series of micro-commands issued by the quantum system controller, specifying the moves and turns of individual qubits and the gate level operations. A complete computational solution is thus obtained as a pair, the initial *placement* $P_1$ of qubits and a *trace* of quantum control micro-commands, which we denote as $T_1$. In addition, as an incidental effect, we produce a new placement $P'_1$ for the outputs qubits (and hence the corresponding input qubits.)

The next step starts by executing each instruction in the UIDG according to S* and using $P_1'$ as the initial placement solution. The result is another control trace $T'_1$ and a new qubit placement, $P_2$. We denote as an *iteration* this sequence composed of a forward computation on QIDG resulting in a placement solution $P_1$, a control trace $T_1$, and a forward latency $L_1$ followed by a backward computation on UIDG resulting in another placement $P_2$, a control trace $T'_1$, and a backward latency $L'_1$. We perform *K* iterations, and in the end, pick the forward or backward computation that results in the smallest latency. If the optimum latency is achieved for a forward computation in the $k^{th}$ iteration, the reported solution is $P_k$, $T_k$, and $L_k$. If, however, the optimum latency is achieved for a backward computation in the $k^{th}$ iteration, the reported solution is $P_{k+1}$, reverse of $T'_k$, and $L'_k$.

Clearly the runtime and solution quality of the proposed placer will depend on *K*. Our proposed placement algorithm makes use of the reversibility of computations in the quantum domain. Also, it takes the scheduling solution into account as opposed to the standard VLSI placement algorithms, which consider only node connectivity (and/or direction of edges) in the given netlist. This is why an iterative approach where the initial placement of a forward computation on the QIDG is the result of the placement obtained from a backward computation on the UIDG and vice versa will find good placement solutions. The quality of our placer has been further improved by starting with *m* random center placements (random seeds) and, for each such seed, doing a local neighborhood search with a variable number of iterations (as described above.) These neighborhood searches are stopped when the quality of the result does not improve for three consecutive placement runs. We call this method as *Multi-start Variable-length Forward/Backward* (MVFB) placer. The sensitivity analysis with respect to *m* is reported in the experimental results section of this paper. We only point out here that the latency of a solution obtained by *m'* placements (which denotes the total number of placements in MVFB with *m* random seeds) based on our proposed algorithm is lower than that of the best *m'* random central placement solutions.

### B. Routing of Two Qubits

All we need to explain at this point is how to route the source and destination operands of a quantum instruction. The process involves (i) selection of the target trap in which the gate level operation between the two qubits will take place, (ii) the routing of the two qubits toward this target trap site.

The target trap site is chosen to be near the median location of the destination and source qubits in the X and Y directions. In particular the median location is first calculated, then a search is conducted to find the nearest available trap to this median location. The goal of the routing step is to minimize $T_{routing} + T_{congestion}$. This is done by creating a weighted graph to model the fabric. In this graph, each vertex represents a junction and each edge represents a channel. The weight of an edge is defined as

$$edge\ weight = \begin{cases} (n+1) \times channel\ length, & n < channel\ capacity \\ \infty, & otherwise \end{cases} \quad (2)$$

where *n* is the number of qubits that are already using or will use the channel as a part of their routing. Increasing weight of an edge discourages the router to select a path containing that edge, and hence, decreases the channel routing and congestion delays. Note that the channel routing delay is proportional to channel length and that *n* captures the congestion in the channel – so the edge weight accounts for both $T_{routing}$ and $T_{congestion}$. If a channel is fully congested, its edge weight goes to infinity to avoid further use of the channel. When a qubit exits a channel, the weight of the corresponding edge will be decreased by *channel length*. This model has an important shortcoming i.e., it does not account for the turn delay, which is a large overhead in the routing. Figure 5.a shows a small circuit fabric example. The mentioned graph model of this fabric is depicted in Figure 5.b. To route from the bottom left corner to the top right corner of this fabric, as indicated in the figure, the router is free to select any of the paths with equal Manhattan distances. Three possible choices are shown in Figure 5.b. However, path (1) is the true shortest path, since it contains only one turn. The effect of turn delay has not been considered in other models in the literature [2] [4][5].

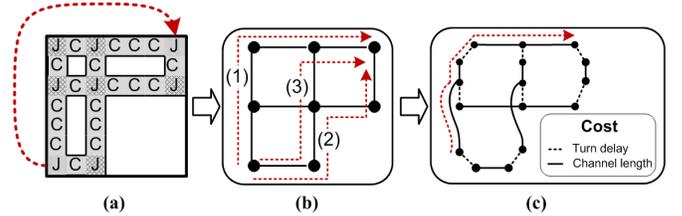

Figure 5. Graph modeling of a fabric for routing. (a) A tile of fabric. The goal is to find the best route from the bottom left corner to the top right corner of the fabric. (b) A weighted simple graph to model the fabric. Three possible paths for routing are shown. Although path (1) is the best path, it looks the same as path (2) and (3) to the router. (c) The enhanced graph model to support turns. The shortest path is now have only one turn.

To support the turn delay, the previous model is changed as follows: Each vertex is replaced with two vertices. The first one connects the horizontal channel(s), whereas the second one connects the vertical channel(s.) These two new vertices are connected to each other through an edge. This edge represents the required turn and its weight is equal to the turn delay in the given technology. Figure 5.c shows the abovementioned enhancement to support the turn delays. The newly added edges are shown by dashed black lines.

When an instruction is taken from the scheduler, the router tries to route it on the fabric graph using the Dijkstra's shortest path algorithm with the edge weighting function described above. If a path is found, it is returned as the solution after the weights of all edges on the path is appropriately increased. If it is impossible to route an instruction due to channel congestion, the instruction is added to a *busy queue*. Instructions will be fetched and re-executed from this queue only after the status of some channels in the circuit fabric changes. Note that an *event driven simulator* is continuously in operation, keeping track status of routing resources, delays of gate level operations, moves and bends. The events are as follows:

- **Execution of an instruction finishes --** The simulator schedules more instruction(s) that depend(s) on the finished instruction.
- **A qubit (from a previously scheduled instruction) exits a channel --** The edge weight of the routing channel is appropriately decreased. The simulator issues more instructions from the busy queue.

## V. EXPERIMENTAL RESULTS

### A. Experimental Setup

QSPR is implemented in Java. The parameters used for the simulations are as follows: $T_{move}$=1μs; $T_{turn}$=10μs; $T_{\text{1-qubit gate operation}}$=10μs; $T_{\text{2-qubit gate operation}}$=100μs; Channel Capacity= 2. A PC with Intel Core i7-2820QM CPU and 8GB DDR3-RAM was used to run the simulations.

An ideal circuit fabric model is defined to evaluate the efficiency of QSPR results. In this model, it is assumed that $T_{congestion}$=0μs and $T_{routing}$=0μs. The execution latency of this ideal model (which we shall call *baseline*) is a lower bound on the execution time of any placed and routed result. A Monte Carlo placer is implemented that places qubits in the nearest traps to the center of the fabric in different permutations. *m'* permutations are randomly selected as initial placements, and the scheduled instructions are routed for each of them. The execution latency of the circuit is derived for each placement. Then, the best result in terms of latency is selected.

Quantum circuits are susceptible to environmental noise, so, encoding circuits play the key role in real quantum circuits. They are located on the critical path of circuit [5], and improving their latency improves the overall latency of the circuit. Hence, to evaluate QSPR, six encoding circuits are selected from [6] and composed in QASM and fed to QSPR as benchmark. Besides, the fabric presented in Figure 4, is used for all of the benchmarks.

### B. Simulation Results

Table 1 shows simulation results for our proposed placer, MVFB, and the Monte Carlo (MC) placer for two cases, *m*=25 and *m*=100. Note that because MVFB uses a variable number of iterations per placement seed (random initial placement solution), we will only know the total number of placements runs after MVFB placer is done. For each testbench circuit, the number of placement runs for the MC approach was set to be exactly twice the number of iterations that MVFB approach used.

Table 1. Comparison of MVFB and MC methods based on execution latency and CPU runtime for m=25 and m=100

| Circuit | Heuristic | m=25 | | | m=100 | | |
|---|---|---|---|---|---|---|---|
| | | Exec. Latency (μs) | CPU Runtime (ms) | Number of Placement Runs | Exec. Latency (μs) | CPU Runtime (ms) | Number of Placement Runs |
| [[5,1,3]] | MVFB | 634 | 546 | 88 | 634 | 921 | 312 |
| | MC | 664 | 562 | | 674 | 967 | |
| [[7,1,3]] | MVFB | 610 | 624 | 78 | 603 | 1154 | 312 |
| | MC | 618 | 640 | | 622 | 1232 | |
| [[9,1,3]] | MVFB | 1159 | 843 | 86 | 1138 | 1529 | 308 |
| | MC | 1212 | 858 | | 1198 | 1576 | |
| [[14,8,3]] | MVFB | 3390 | 1638 | 83 | 3342 | 3666 | 316 |
| | MC | 3540 | 1654 | | 3429 | 4087 | |
| [[19,1,7]] | MVFB | 3393 | 2387 | 82 | 3350 | 6598 | 311 |
| | MC | 3483 | 2699 | | 3403 | 7098 | |
| [[23,1,7]] | MVFB | 2066 | 3942 | 89 | 2061 | 9298 | 315 |
| | MC | 2183 | 4527 | | 2085 | 10905 | |

As can be seen, the MVFB placer produces higher quality results (i.e., lower execution latency for the quantum circuit mapped to the ion-trap circuit fabric) for both m=25 and m=100 cases. By design of experiment, the CPU runtimes of the MVFB and MC placers are nearly the same, with the MVFB placer being a bit faster on larger circuits due to doing local neighborhood searches on a smaller number of random seeds (although the total number of placement runs for both cases are the same.) Note also that, as expected, the m=100 case always yields better results than m=25 for both MVFB and MC placers.

Table 2 compares the (ideal) baseline, QUALE, and QSPR in terms of the execution latency of the mapped (scheduled, placed and routed) circuits. Note that QUALE is the only publically available quantum circuit mapping tool. For QSPR, the MVFB placer is chosen with *m*=100. First, note that QSPR's results are superior to those of QUALE. The last column shows percentage improvement of QSPR over QUALE. The improvement is from 24% up to 55%. This value increases as the latency of the base circuit increases. Second, the latency difference between the baseline results and the QSPR/QUALE results shows the impact of $T_{routing}$+$T_{congestion}$. The general trend shows that $T_{routing}$+$T_{congestion}$ have higher impact on the latency of larger circuits.

Table 2. Comparison of Baseline, QUALE and QSPR in terms of execution latency of QECC circuits.

| Circuit | Heuristic | Execution Latency (μs) | Difference wrt Baseline (μs) | Improvement wrt QUALE (%) |
|---|---|---|---|---|
| [[5,1,3]] | Baseline | 510 | - | 23.80 |
| | QUALE | 832 | 322 | |
| | QSPR | 634 | 124 | |
| [[7,1,3]] | Baseline | 510 | - | 23.56 |
| | QUALE | 798 | 288 | |
| | QSPR | 610 | 100 | |
| [[9,1,3]] | Baseline | 910 | - | 47.70 |
| | QUALE | 2216 | 1306 | |
| | QSPR | 1159 | 249 | |
| [[14,8,3]] | Baseline | 2500 | - | 54.87 |
| | QUALE | 7511 | 5011 | |
| | QSPR | 3390 | 890 | |
| [[19,1,7]] | Baseline | 2510 | - | 50.38 |
| | QUALE | 6838 | 4328 | |
| | QSPR | 3393 | 883 | |
| [[23,1,7]] | Baseline | 1410 | - | 44.73 |
| | QUALE | 3738 | 2328 | |
| | QSPR | 2066 | 656 | |

## VI. CONCLUSION

In this paper, we targeted the circuit latency as the objective function and reduced it to minimize the amount of noise a quantum circuit absorbs. This was done by building a CAD tool, which improved the mapping of a QASM file on a given ion-trap fabric by 24% to 55% wrt the previous tool. A heuristic called MVFB placement was proposed which improved the center placement used in the previous tool. This heuristic was shown to generate higher quality results than the Monte Carlo method. Moreover, it was observed that $T_{routing}$ and $T_{congestion}$ play an important role in the latency of larger circuits.

### ACKNOWLEDGMENT

The authors would like to thank Todd Brun for his valuable comments on the ion-trap fabric and recent advances in the quantum computing technologies.